\documentclass[aps,superscriptaddress,showpacs,nofootinbib,floatfix, twocolumn]{revtex4}
\usepackage{graphicx}
\usepackage{bm,color,ulem, soul}
\usepackage{amssymb}
\usepackage{CJK}
\newcommand{\me}{\varepsilon}

\newcommand{\dd}{\textrm{d}}

\newcommand{\erf}{\textrm{erf}}

\newcommand{\bc}[1]{[{\bf \color{blue}{baoyichen}}]}

\begin{document}
\author{Guangyu Li}
\author{Baoyi Chen}
\author{Yunpeng Liu}
\email{Email:yunpeng.liu@tju.edu.cn}
\affiliation{Department of Applied Physics, Tianjin University, Tianjin 300350, China}
\pacs{25.75.Nq, 12.38.Mh, 25.75.-q}
% 12.38.Mh	QGP
% 25.75.Nq	Quark deconfinement, quark-gluon plasma production, and phase transitions 
% 64.60.an	
% 12.39.Ba	Bag model
% 25.75.-q	Relativistic heavy ion collisions
% 14.40.Pq	Heavy quarkonia
\title{Relativistic correction to the dissociation temperature of $B_c$ mesons in the  hot medium}
\begin{abstract}
By solving two body Dirac equations with potentials at finite temperature, we calculated the dissociation temperature $T_d$ of $B_c$ mesons in the quark-gluon plasma. 
It is found that the $T_d$\ becomes higher with the relativistic correction than the $T_d$ from the Schr\"odinger equation. Both the short range interaction and the constant term of the potential at the long-range scale have a contribution to the shift of $T_d$, while the spin interaction is negligible. 
\end{abstract}
\maketitle
\section{Introduction}
The phase transition between the quark-gluon plasma(QGP) and the hadron phase is one of the most interesting topics in the relativistic heavy ion physics. As the QGP can not be detected directly, the heavy quarkonium consisting of a heavy quark and a heavy antiquark has been proposed as a probe for the early stage of the hot medium in heavy-ion collisions~\cite{Matsui:1986dk, Xu:1995eb, Thews:2000rj, Zhuang:2003fu, Yan:2006ve, Zhao:2007hh, Brambilla:2010vq, Song:2011xi, Du:2019tjf, Deng:2016stx, Guo:2015nsa}. Non-relativistic approximation is usually taken in the studies of heavy quarkonium, due to its large mass~\cite{Satz:2005hx, Liu:2009nb, Brambilla:2010vq, Chen:2018dqg, Blaschke:2005jg}. Such an approximation is usually reasonable at zero temperature when the mass or radius is calculated, because the typical energy scale of the potential is smaller than the masses of heavy quarks. 

However, the dissociation with a screened potential is different. The temperature as a new low energy scale comes in. The potential pulls the quarks together into a meson, and the temperature weakens the attraction.  Around the maximum survival temperature, that is the  dissociation temperature $T_d$, the balance between the attractive potential and the screening can be destroyed even by a relatively small contribution in the potential.  Therefore, it is interesting to check how much change in temperature can rebalance the relativistic correction. Determining the in-medium properties of quarkonium is crucial for extracting the heavy quark potential at finite temperatures and also the hot medium information with the quarkonium probes.  Some theoretical studies~\cite{Guo:2012hx, Shi:2013rga, Wu:2012ns} about the relativistic effect have been done in open and hidden heavy flavors based on two body Dirac equations (TBDE)~\cite{Crater:1987hm, Long:1998, Liu:2002cn, Crater:2012ih}. 

An analog to heavy quarkonia is the $B_c$ meson~\cite{Schroedter:2000ek, Akram:2012zbm, Liu:2012tn, Alberico:2013pha, Chen:2021uar}, since it is composed of two heavy quarks. It is even a cleaner probe than heavy quarkonia, because (i) it is composed of heavy quarks with different flavor, so that the possible processes are less than heavy quarkonia, and (ii) $B_c$ is a pseudo-scalar meson, so that the spin structure is simpler. 
It is therefore urgent to study the properties of the $B_c$ meson at finite temperatures. In experiments, the nuclear modification factor of $B_c$ mesons has been measured, and more helpful results can be expected in the future~\cite{CMS:2022sxl}. In this work, we will employ TBDE compared with the Schr\"odinger equation to study the relativistic correction on $B_c$ dissociation temperature. 

\section{TBDE for $B_c$ mesons at finite temperature}
The TBDE is a generalization of the free Dirac equation. It is both relativistic and Schr\"odinger-like, and it has been used to calculate the meson spectra in vacuum and at finite temperature. In general, it is complicated coupled equations~\cite{Crater:2012ih, Guo:2012hx}. For a (pseudo-)scalar meson, it decouples from the spin triplet components, and its equation can greatly be simplified. We rewrite its radial equation as
\begin{eqnarray}
	\left[-\frac{\dd^2}{\dd r^2}+\Phi_{SI}+\Phi_1\right]U(r)&=&b^2 U(r),
\end{eqnarray}
with
\begin{eqnarray}
	\Phi_{SI}&=&2m_wS+2\varepsilon_w A+S^2-A^2,\\
	\Phi_{1}&=&\frac{1}{R}\frac{\dd^2}{\dd r^2}R,\\
	R&=&r\sqrt{\frac{(m_1\me_2+m_2\me_1)X}{\me_1M_2+\me_2M_1-A(M_1+M_2)}},\\
	M_i&=&\sqrt{m_i^2+Y/X},\quad (i=1,2)\\
	Y&=&2m_wS+S^2,\\
	X&=&1-\frac{2A}{m_m}.
\end{eqnarray}
Here $r$ is the radius of the relative motion. The quantities denoted by capital letters are all $r$-dependent functions. This convention is also kept elsewhere in this paper except for conventional special functions, the temperature $T$, and $r$ itself. The $r$-independent quantities are the constituent mass of each quark $\me_i$, the reduced mass $m_w$, the energy $\me_w$ of the imaginary particle for relative motion, and the eigenvalue-like term $b^2$, which are defined as
\begin{eqnarray}
	\me_i&=&\frac{m_m^2+2m_i^2-m_1^2-m_2^2}{2m_m}, (i=1,2)\\
	m_w&=&\frac{m_1m_2}{m_m},\\
	\me_w&=&\frac{m_m^2-m_1^2-m_2^2}{2m_m},\\
	b^2&=&\frac{m_m^4-2(m_1^2+m_2^2)m_m^2+(m_1^2-m_2^2)^2}{4m_m^2}.
\end{eqnarray}
The inputs of such an equation are the charm quark mass $m_1$, the bottom quark mass $m_2$, and the potential between them divided into a scalar part $S$ and a vector part $A$, while the outputs are both the mass $m_m$ of the $B_c$\ meson and its radial wave function $U(r)$, so that the probability to find the quark pairs in a sphere shell $r\sim r+\dd r$ is proportional to $|U^2(r)|\dd r$. Note that $m_m$ appears on both sides of the equation. Therefore, it is actually not a typical eigen equation.

In the non-relativistic limit $A, S\ll m_1, m_2, m_w, m_m$.\footnote{Rigorously, both $A$ and $S$ are functions of $r$. This condition means that in the typical range of $r$ where $|U(r)|^2$ is reasonable large, $A$ and $S$ are far smaller than $m_1, m_2, m_w$ and $m_m$.} one find $X\rightarrow 1$, $M_i\rightarrow m_i$, $R\rightarrow r$, $\Phi_1\rightarrow 0$, $\me_w\rightarrow m^S$, $m_w\rightarrow m^S$ with the non-relativistic reduced mass  $m^S=\frac{m_1m_2}{m_1+m_2}$, the spin-independent term $\Phi_{SI}\rightarrow 2m^SV^S$ with $V^S=A+S$ being the non-relativistic potential, and $b^2\rightarrow 2m^S\me$, with $\me=m_m-m_1-m_2$ being the eigen energy in a Schr\"odinger equation. As expected, it reduces to a Schr\"odinger equation
\begin{eqnarray}
   \left[-\frac{1}{2m^S}\frac{\dd^2 }{\dd r^2}+V^S \right]U(r)&=&\me U(r),
\end{eqnarray}
or
\begin{eqnarray}
   \left[-\frac{1}{2m^S}\frac{\dd^2 }{\dd r^2}+\overline{V}^S\right]U(r)&=&-\Delta \me^S U(r),
\end{eqnarray}
with the non-relativistic binding energy $\Delta \me^S= m_1+m_2+v_{\infty}-m_m$. Here we have introduced the shifted potential $\overline{V}^S=V^S-v_{\infty}$, $v_{\infty}=\lim\limits_{r\rightarrow +\infty}V^S$, so that $\overline{V}^S$ vanish at $r\rightarrow +\infty$. Such a bar-symbol is also used for other potentials in the following.

A widely used potential $V^S$ in vacuum is the Cornell potential which contains a Coulomb-like potential $A=-\frac{\alpha}{r}$, and a linear potential $S=\sigma r$. Charmonium and Bottomonium spectra can be fitted with $\alpha=\pi/12$ and $\sigma=0.2$\ GeV$^2$~\cite{Satz:2005hx}. At finite temperature, we take the screened Cornell potential~\cite{Satz:2005hx} as a function of the screening mass $\mu$ and radius $r$, 
\begin{eqnarray}
	V^S(r,\mu)&=&A(r,\mu)+S(r,\mu),\\
	A(r,\mu)&=&-\alpha\mu\left(\frac{e^{-\mu r}}{\mu r}+1\right),\\
	S(r,\mu)&=&\frac{\sigma}{\mu\Gamma(\frac{3}{4})}\left[ \frac{\Gamma(\frac{1}{4})}{2^{\frac{3}{2}}}-\frac{(\mu r)^{\frac{1}{2}}}{2^{\frac{3}{4}}}K_{\frac{1}{4}}(\mu^2 r^2) \right],
\end{eqnarray}
We have attributed the screened Coulomb-like potential to $A$, and the screened confining potential to $S$, as it is at zero temperature~\cite{Crater:2008rt}. This is our assumption in the main part of this paper, and we will revisit it in the end.
The only temperature dependent parameter is the screening mass $\mu$, which we fit from the lattice result~\cite{Digal:2005ht} as 
\begin{eqnarray}
\mu(\overline{T})/\sqrt{\sigma}&=& s\overline{T}+\sqrt{\frac{\pi}{2}}a\sigma_{t}\left[\erf(\frac{b}{\sqrt{2}\sigma_{t}})-\erf(\frac{b-\overline{T}}{\sqrt{2}\sigma_t})\right],\nonumber\\
\end{eqnarray}
with $\overline{T}=T/T_c$, $\erf(t)=\frac{2}{\sqrt{\pi}}\int_{0}^{t}e^{-x^2}\dd x$,
$s=0.587$,$a=2.150$, $b=1.054$, and $\sigma_t=0.07379$.

The only free parameters left are quark masses $m_1$ and $m_2$. 
We fit masses of scalar heavy mesons composed of a charm quark and/or a bottom quark in experiments~\cite{PDG:2022} to obtain $m_1$ and $m_2$ both in TBDE and in the Schr\"odinger equation. The results are listed in Table~\ref{tb_mq}.
We denote the results or parameters for the Schr\"odinger equation with a superscript $^S$, and those for TBDE with a superscript $^D$ to distinguish them when necessary.
\begin{table}[!hbt]
\begin{tabular}{|l|ll|lll|}
	\hline
	&$m_1$ & $m_2$ & $m_{\eta_c}$ & $m_{\eta_b}$ & $m_{B_c}$\\
	\hline
	Schr\"odinger Eq.& 1.19 & 4.59 &  2.991 & 9.415 & 6.256\\
	\hline
	TBDE&1.27&4.62&2.997&9.421&6.254\\
	\hline
	Experiments&-&-&2.984&9.399&6.274\\
	\hline
\end{tabular}
\caption{Mass parameters in the Schr\"odinger equation and those in TBDE. The units of masses in the table are all GeV. Experimental data are from the Particle Data Group~\cite{PDG:2022}.}
\label{tb_mq}
\end{table}

\section{Results and discussions}
\begin{figure}[!hbt]
    \centering
    \includegraphics[width=0.4\textwidth]{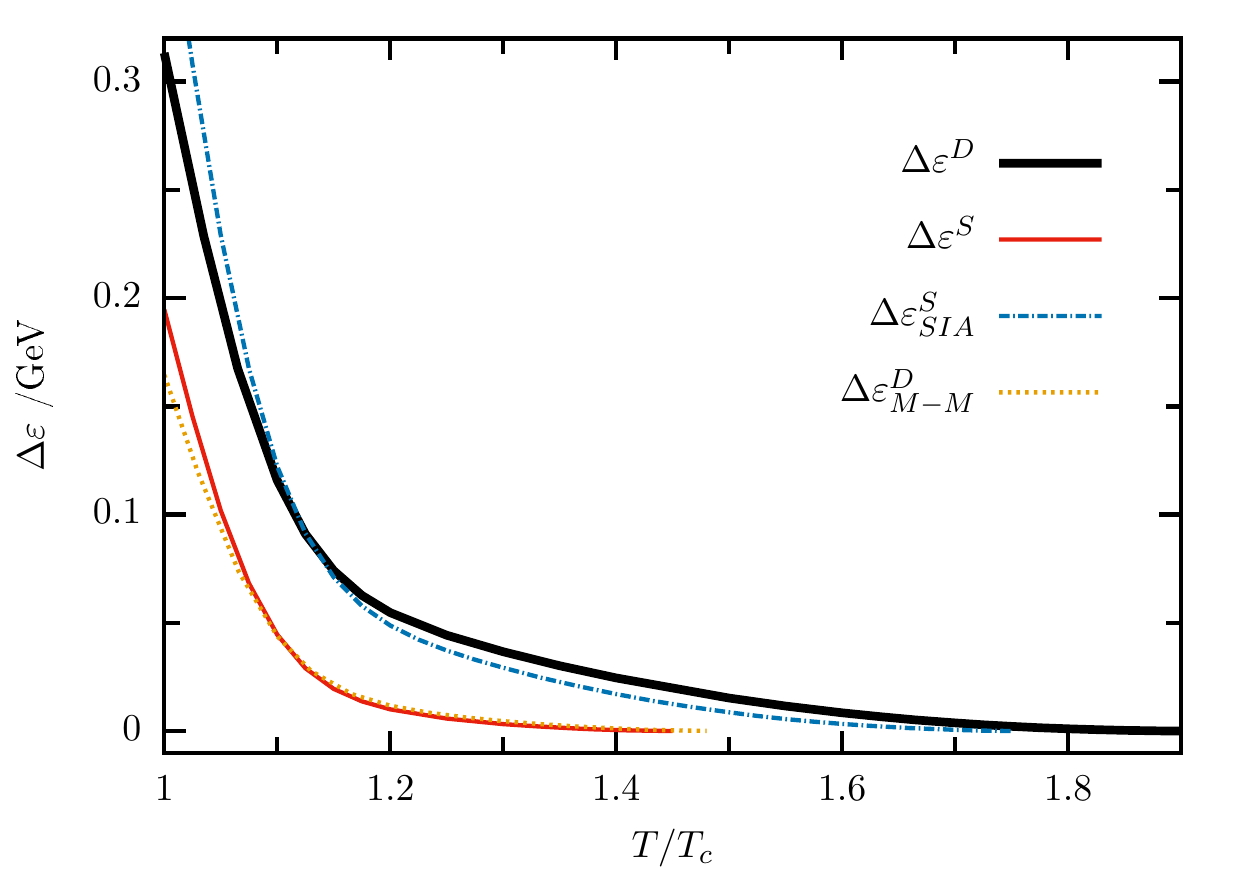}
    \caption{Binding energy $\Delta \me$ as a function of temperature $T$ compared with the non-relativistic limit.}
    \label{fg_epsilon_T}
\end{figure}

\begin{figure}[!hbt]
    \centering
    \includegraphics[width=0.4\textwidth]{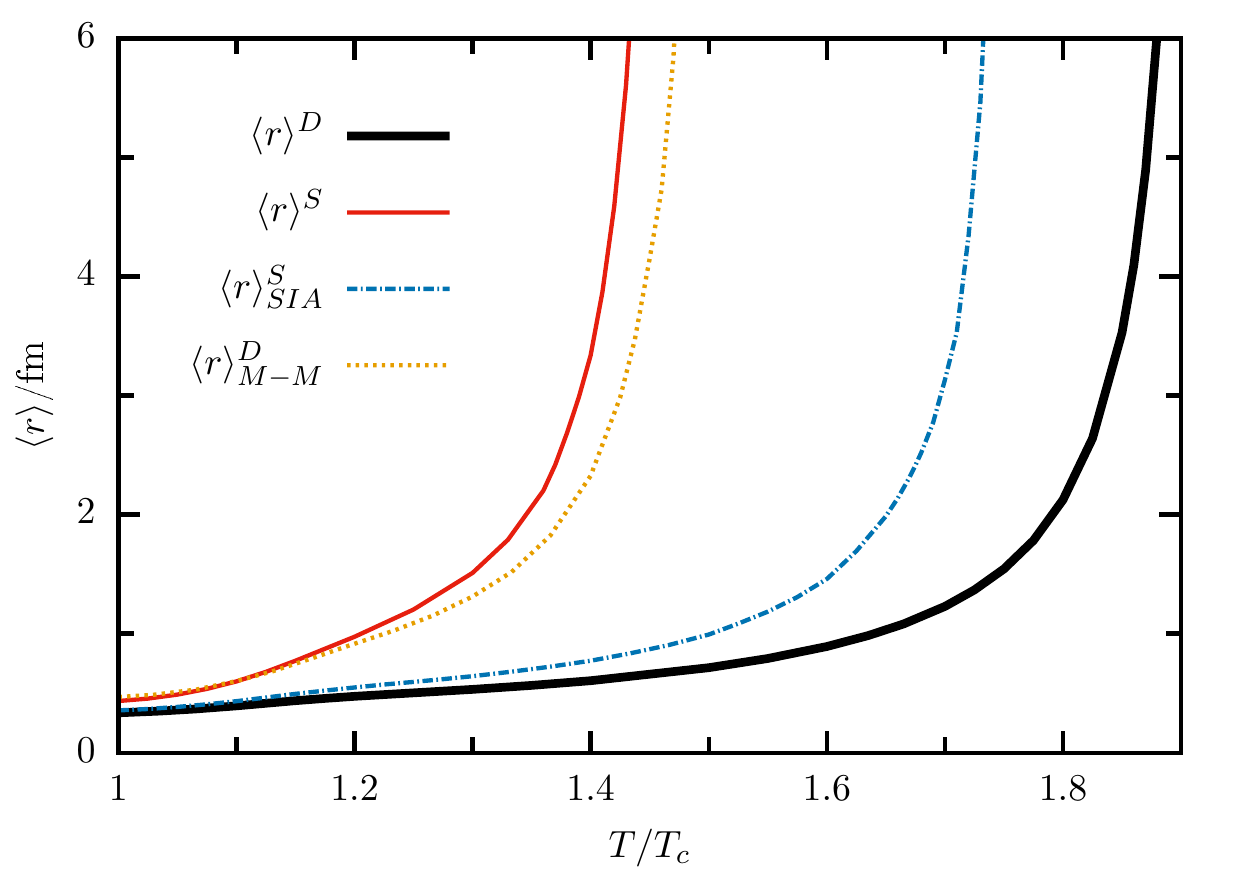}
    \caption{Average radius as a function of temperature $T$ compared with the non-relativistic limit.}
    \label{fg_r_T}
\end{figure}

\begin{figure}[!hbt]
    \centering
    \includegraphics[width=0.4\textwidth]{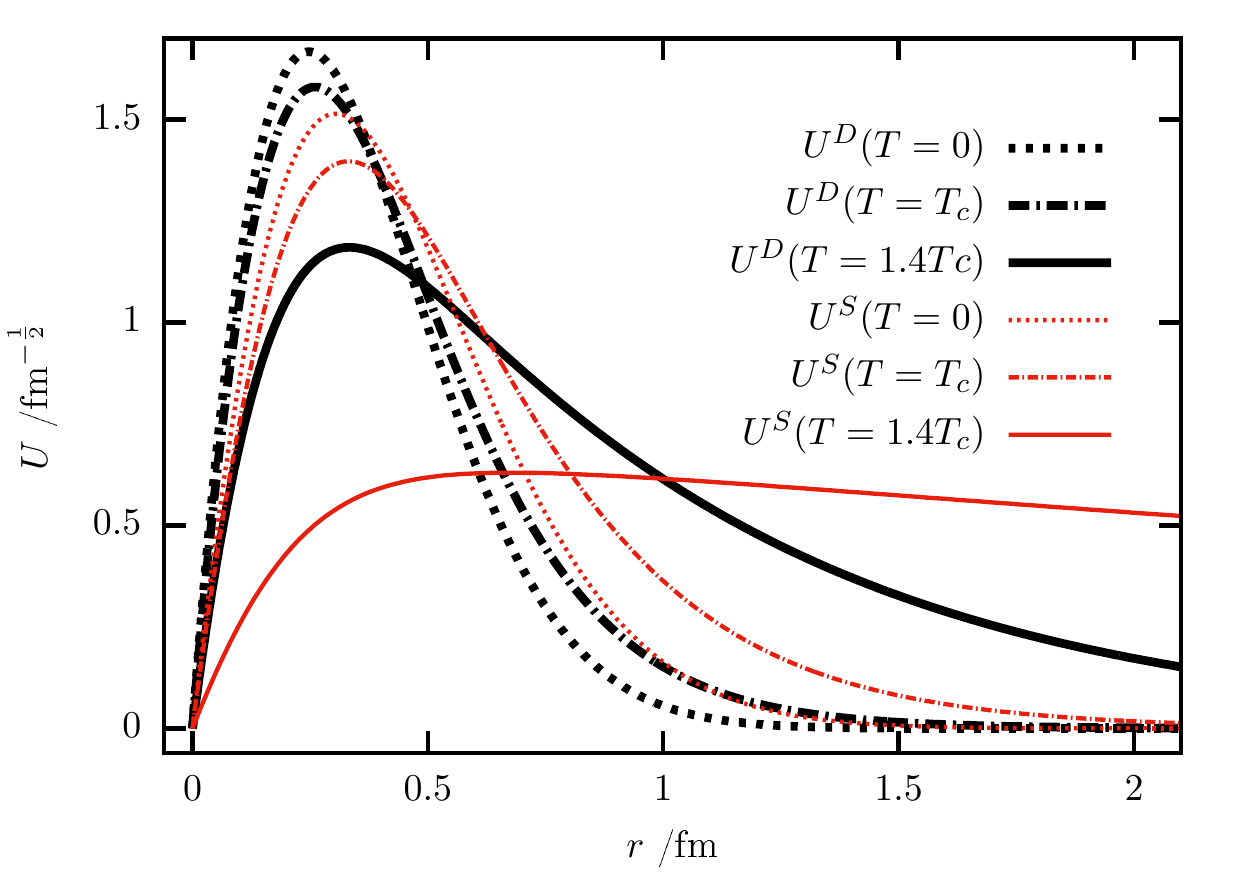}
    \caption{Wave functions at different temperatures $T=0$, $T=T_c$ and $T=1.4T_c$.}
    \label{fg_wave_function}
\end{figure}

At finite temperature, the binding energy $\Delta \me$ of $B_c$ is defined  as the energy difference between the lowest scattering state and the bound state
\begin{eqnarray}
   \Delta \me^D &=& v_{\infty}+\sqrt{v^2_{\infty}+(m_1+m_2)^2}-m_m^D,
\end{eqnarray}
while in the non-relativistic limit, it becomes $\Delta \me^S = v_{\infty}+m_1+m_2-m_m^S$ as mentioned above. The temperature dependence of the binding energy $\Delta \me$, and that of the average radius
\begin{eqnarray}
	\langle r \rangle&=& \frac{\int \dd r\ r |U(r)|^2}{\int \dd r |U(r)|^2},
\end{eqnarray}
and the $r$ dependence of the wave functions $U$ of $B_c$ are shown in Fig.~\ref{fg_epsilon_T}, Fig.~\ref{fg_r_T}, and Fig.~\ref{fg_wave_function}, respectively.
It can be seen that in both cases, the higher the temperature is,  the weaker the binding between the two heavy quarks is, as expected. The dissociation temperature $T_d$, at which the binding energy reduces to zero, increases from $1.45T_c$ to $1.9T_c$ when the relativistic correction is considered. 

To understand the differences, we firstly check the parameters $m_1$ and $m_2$ dependence. It is found that this factor is less important. For example, if the TBDE parameters $m_1$ and $m_2$(see the middle row in Table~\ref{tb_mq}) instead  are used in the Schr\"odinger equation, the dissociation temperature $T_d$ will increases only by about only $0.05T_c$.
This indicates that the differences in $T_d$ are mainly due to the relativistic corrections in the equations, instead of the parameter difference in Table 1.

Now we turn to the difference in equations. A direct comparison is to compare $\Phi$\footnote{Here we substitute the value of solved $m_m$ into $\Phi$ so that TBDE can be treated as a typical eigen equation, and thus it can be compared with the Schr\"odinger equation.}. However, a potential is more intuitive. Therefore, we consider potentials $V$ with and without the relativistic corrections at different temperatures $T_c$ and $1.4T_c$, with $V^D\equiv \frac{1}{2m^S}\left( \Phi_{SI}+\Phi_1 \right)$. 
%For a better comparison, we plot $\overline{V}$ in practice in Fig.~\ref{fg_VT} instead of $V$. That is we have dropped the finite constant term at $r\rightarrow +\infty$ in the potential.
\begin{figure}[!hbt]
    \centering
    \includegraphics[width=0.4\textwidth]{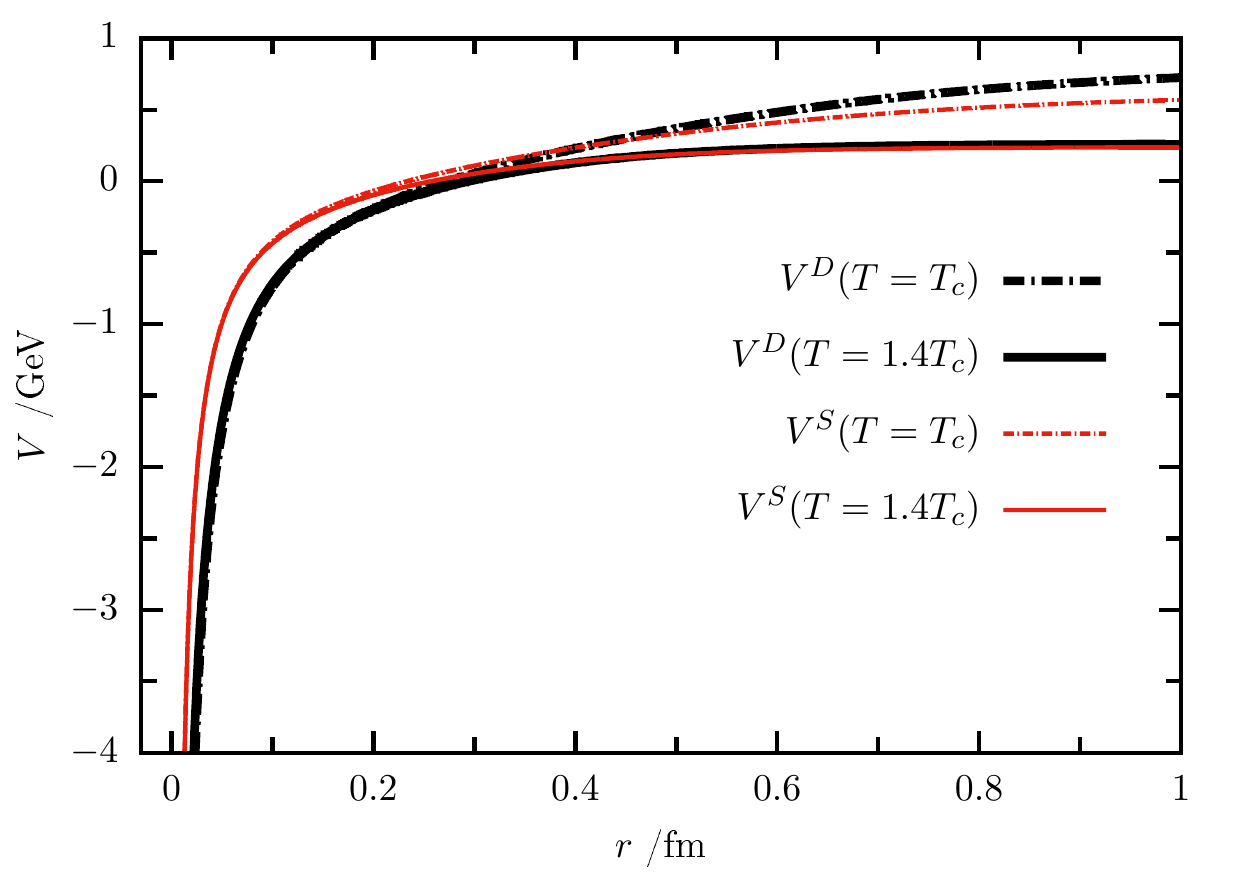}
    \caption{Potentials as functions of radius $r$ at temperatures $T=T_c$\ and $T=1.4T_c$.}
    \label{fg_VT}
\end{figure}
It can be seen that at $r\rightarrow 0$, the temperature dependence of ${V}$ is weak, because at short distance it is dominated by the Coulomb term of the potential $A$, and the screening is weak. The effective potential $V^D$ in TBDE is deeper than the potential $V^S$ in the Schr\"odinger equation at short range mainly because of the $-A^2\approx-\alpha^2/r^2$ term.  At large $r$, the difference between $V^D$ and $V^S$ are much smaller, which is reasonable, since a large radius corresponds to a low energy scale, and the relativistic effect is less important. The differences are from the constant term in the screening potentials. In short, the potential well becomes deeper at lower temperature and/or in the relativistic case.

%It can be seen that at $r\rightarrow 0$, the temperature dependence of $\overline{V}$ is relatively weak because at short distance it is dominated by the Coulomb term and the screening of it is weak, while at $r\rightarrow +\infty$, we have $\overline{V}^S\approx\overline{V}^D$, which is reasonable since a large radius corresponds to a low energy scale, and the relativistic effect is less important. The potential well becomes deeper at lower temperature and/or in the relativistic case.

\begin{figure}[!hbt]
    \centering
    \includegraphics[width=0.4\textwidth]{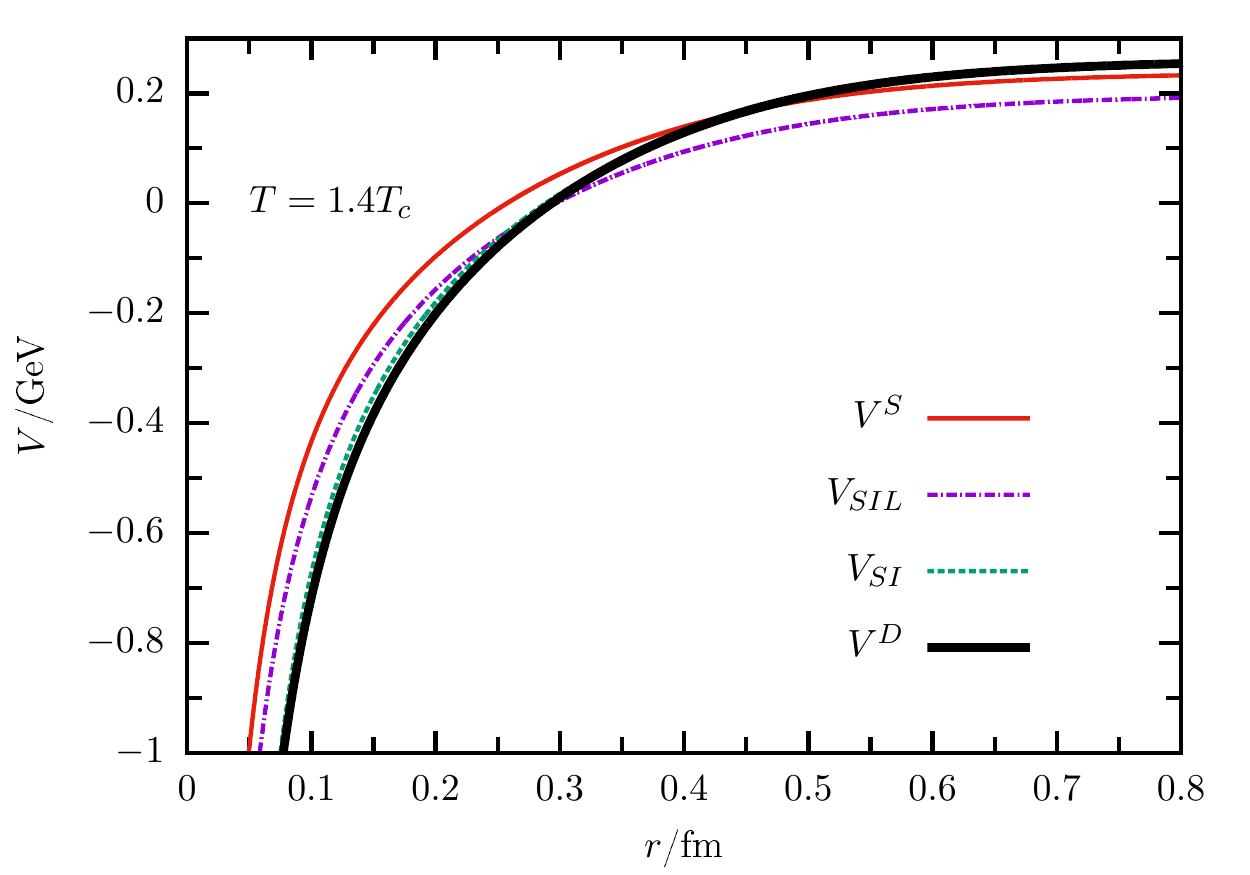}
    \caption{Potentials as functions of radius $r$ at temperature $T=1.4T_c$.}
    \label{fg_V}
\end{figure}
The potential term is dominated by $\Phi_{SI}$, which means that the dissociation temperature of $B_c$ meson is not sensitive to the spin structure. Actually, up-to the next-to-leading order of the non-relativistic approximation, we have
\begin{eqnarray}
   R&=& r\left[1+\frac{\sec^2\theta}{m_m}A-\frac{\sec^2\theta+\tan^2\theta}{m_m}S\right],
\end{eqnarray}
with $\sin\theta=(m_1-m_2)/m_m$. The leading term of $\Phi_1$\ is
\begin{eqnarray}
   \Phi_1
   &=& \frac{1}{m_mr}\left[\sec^2\theta (rA)''-(\sec^2\theta+\tan^2\theta) (rS)''\right]\nonumber\\
&\sim&\frac{\mu^2}{m_m}\overline{V}^S,
\end{eqnarray}
which implies that $\Phi_1$\ is roughly smaller than $m_mV^S$ by a factor of $\frac{\mu^2}{m_m^2}$, and thus roughly smaller than $\Phi_{SI}\approx 2m_wV^S$ by a factor of $\frac{\mu^2}{2m_1m_2}$. At the temperature that the $B_c$ meson dissociates the screening mass is roughly $0.6$ GeV, so that $\Phi_1$ is only as small as few percents of $\Phi_{SI}$, and thus can be neglected, as shown in Fig.~\ref{fg_V} with $V_{SI}=\Phi_{SI}/(2m^S)$\ and $V^{D}=(\Phi_{SI}+\Phi_1)/(2m^S)$.
In this case, TBDE deduce to a relativistic two body equation for spinless particles~\cite{Crater:2012ih}.

The leading term in $\Phi_{SI}$\ is $\Phi_{SIL}=2m_wS+2\me_wA$, which differs from the non-relativistic potential in the mass terms. The masses at $T=T_c$ and $T=1.4T_c$ are listed in Table~\ref{tb_2}. It can be seen that the modification to the scalar part of the potential is relatively small (a few percents), while that to the vector part is relatively large ($\me_w/m^S-1\approx 0.3$). As a result, the modification due the the mass coefficient is roughly at the order of $15\%$. (See the curve of  ${V}_{SIL}=\Phi_{SIL}/(2m^S)$ in Fig.~\ref{fg_V}.)
\begin{table}[!hbt]
   \begin{tabular}{cccccc}
	\hline
	$T/T_c$ & $m_m$	&$m^S$ & $m_w$ & $\me_w$\\
	\hline
	1.0	&	6.217	&0.945&	0.944	&1.262\\
	\hline
	1.4	&	6.111	&0.945&	0.960&1.177\\
	\hline
   \end{tabular}
   \caption{Masses in  $\Phi_{SIL}$. All the units of masses in the table are GeV.}\label{tb_2}
\end{table}

Note that there is some ambiguity from the non relativistic potential $V^S$ to the potential $\Phi_{SI}+\Phi_1$ in TBDE, because it depends on how it is divided into the vector part $A$ and the scalar part $S$, especially when the potential is given numerically as in the lattice QCD. We tried to attribute both the $v_{\infty}$ terms to the scalar potential, leaving $A=-\alpha e^{-\mu r}/r$. Then the dissociation temperature shifts from $1.9T_c$ to $1.7T_c$. Actually, we can rewrite $\Phi_{SI}$ under the approximation $m_w\approx m^{S}$ and $\me_w\approx m^S$ as $\Phi_{SI}\approx \Phi_{SIA}=2m^SV^{S}(1+(S-A)/(2m^S))$. Therefore, the larger $S-A$ is, the deeper the well is. In the trial scheme in this paragraph, $S-A$ became smaller, and the binding is looser.
The results from the Schr\"odinger equation with the above $\Phi_{SIA}$ are shown by the blue dotted-dashed curves in Figs.~\ref{fg_epsilon_T} and \ref{fg_r_T}. The results are closer to the TBDE results as expected. Another way one may be interested in to approximate the interaction of quarks is to attribute the constant potential to quark masses~\cite{Riek:2010fk, Liu:2017qah}, i.e. using $m_b'=m_b+v_{\infty}/2$ and $m_c'=m_c+v_{\infty}/2$ to absorb the constant term $v_{\infty}$ in the potentials. The results are shown by the yellow dotted curves in Figs.~\ref{fg_epsilon_T} and \ref{fg_r_T}. In this case the dissociation temperature is slightly higher than the Schr\"odinger results.

\section{Summary}
In summary, the dissociation temperature of $B_c$\ meson in the QGP is calculated by TBDE, and it is found to be higher than that by Schr\"odinger equation. The spin dependent part is negligible, and the increase is mainly due to the square terms $S^2-A^2$ in the spin independent part of the potential $\Phi_{SI}$. In the case $S$ is finite and $A$ approaches to $-\alpha/r$ around $r=0$, the potential well becomes deeper at small $r$. The constant term in the potential which is not important in the Schr\"odinger equation turns to deserve serious consideration in TBDE.

\section*{Acknowledgment} This work is supported by the National Natural Science Foundation of China (NSFC) under Grant No. 12175165. Liu is grateful to Prof. Xingyu Guo for helpful discussions.

\bibliographystyle{unsrt}
\bibliography{refs}
\end{document}